\documentclass{clsf}
\usepackage[utf8]{inputenc}
\usepackage[english,russian]{babel}
\twocolumn
\rus

\usepackage{xcolor}
\usepackage{latexsym,amsmath,amssymb,amsbsy,graphicx}
\usepackage{epstopdf}
\usepackage{epsfig}

\title{Scaling Relations for Temperature Dependences \\of the Surface Self-Diffusion Coefficient\\ in Crystallized Molecular Glasses}

\rtitle{Scaling Relations for Temperature Dependences\ldots}

\sodtitle{Scaling Relations for Temperature Dependences of the Surface Self-Diffusion Coefficient in Crystallized Molecular Glasses}

\author{A.\,V.\,Mokshin$^{a,b,}$\/\thanks[1]{anatolii.mokshin@mail.ru}, B.\,N.\,Galimzyanov$^{a,b,}$\thanks[2]{bulatgnmail@gmail.com}, D.\,T.\,Yarullin$^{a,}$\/\thanks[3]{YarullinDT@gmail.com}}
\address{$^{a}$Kazan Federal University, Kazan, $420008$ Russia}
\address{$^{b}$Udmurt Research Center, Ural Branch, Russian Academy of Sciences, Izhevsk, $426067$ Russia}

\rauthor{A.\,V.\,Mokshin, B.\,N.\,Galimzyanov, D.\,T.\,Yarullin}
\sodauthor{Mokshin, Galimzyanov, Yarullin}

\date{\today}

\abstract{Crystallization kinetics has features that are universal and independent of the type of crystallized system. The possibility of using scaling relations to describe the temperature dependences of the surface self-diffusion coefficient $D_s$, which is one of the key characteristics of crystallization kinetics, has been demonstrated in application to various crystallized molecular glasses. It has been shown that the surface self-diffusion coefficient $D_s$ as a function of the dimensionless temperature is reproduced by a power law and is universally scaled for all considered systems. The analysis of experimental data has revealed a correlation between the crystallization kinetic characteristics, index of fragility, and criterion of the glass-forming ability of a liquid. It has been shown that this correlation can be obtained within the generalized Einstein-Stokes relation.}

\begin{document}
	
\maketitle
The feature of crystallization of molecular glasses such as ortho-terphenyl, griseofulvin, and indomethacin is that the formation of a crystal phase in these systems begins in a surface layer near the molecular glass-air interface~\cite{Zhang_Brian_2015}. This makes it possible to directly observe and detect events associated with the initial stage of crystallization; for this reason, these systems are appropriate candidates for the experimental study of crystallization mechanisms at the atomic/molecular level. It becomes possible to obtain complete information on the key characteristics of processes of nucleation and growth of crystals such as the critical size and shape of formed crystal nuclei, nucleation rate, and growth rate. It is remarkable that these parameters significantly depend on the mobility of particles in the regions of a system where events associated with crystal nucleation occur~\cite{Kashchiev_2000}. For this reason, experimental data on the surface self-diffusion coefficient $D_s$ are of great interest for crystallized molecular glasses.

\begin{figure*}[h!]
	\centering
	\includegraphics[width=0.8\linewidth]{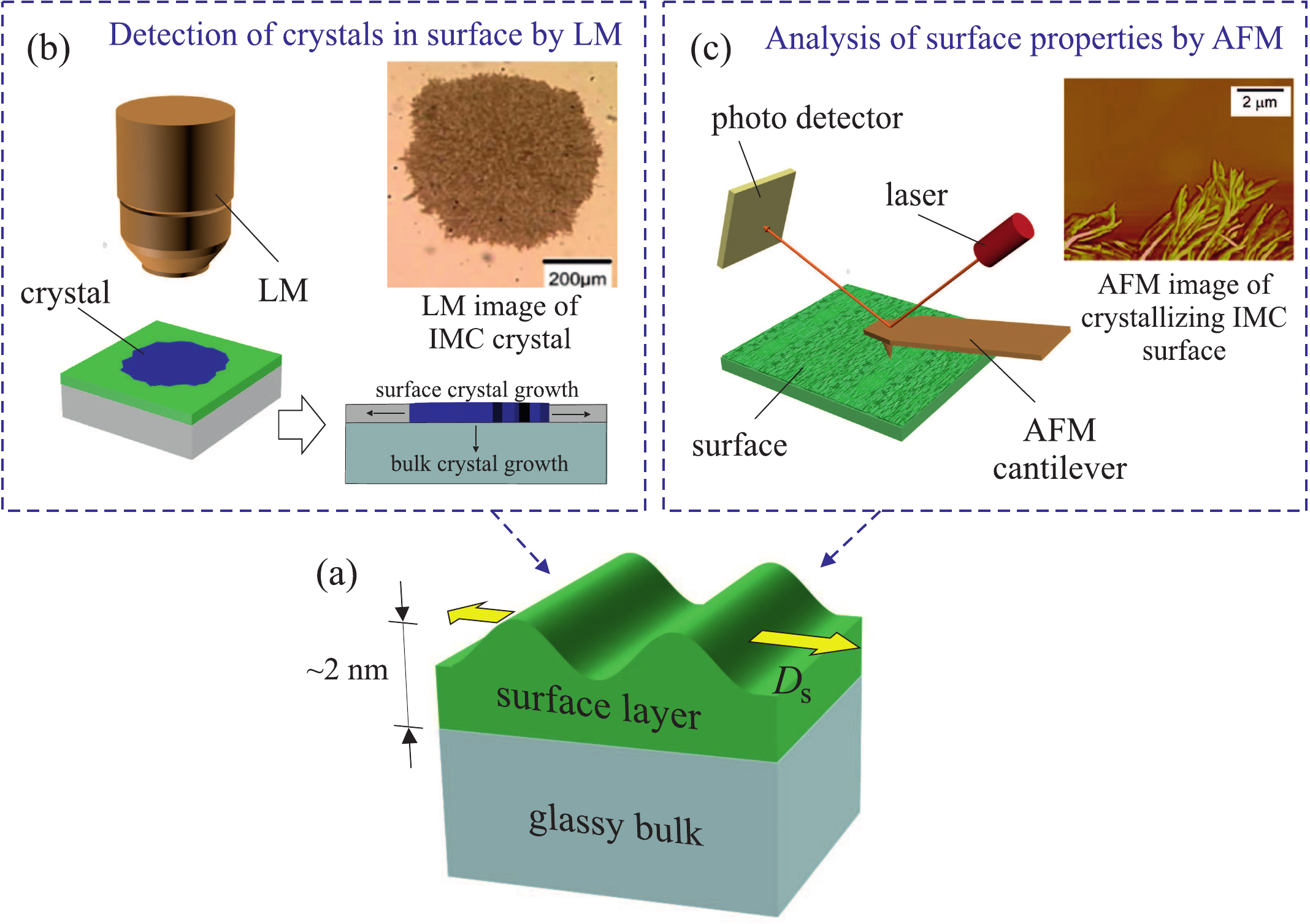}
	\caption{Fig. 1. (Color online) (a) Schematic of the measurement of surface self-diffusion coefficient $D_s$ of crystallized molecular glass (according to~\cite{Sun_2012} and~\cite{Hasebe_2014}). (b) Optical and (c) atomic force microscopy images of crystalline indomethacin (IMC) and nifedipine (NIF) obtained by an optical microscope (LM) and atomic force microscope (AFM) (taken from \cite{Sun_2012} and ~\cite{Hasebe_2014}). The optical microscope can be used to detect crystallites and to analyze their shape. The atomic force microscope is used to detect the relaxation of the deformed surface of an amorphous system and to detect the surface self-diffusion coefficient $D_s$. The atomic force microscopy data are used to calculate the surface self-diffusion coefficient by the Mullins relation~\cite{Mullins_1059}: $D_{s}\propto Kk_{B}T/\gamma_{s}$, where $K$ is the average surface smoothing rate, $\gamma_{s}$ is the surface tension coefficient, $T$ is the temperature of the sample, and $k_{B}$ is the Boltzmann constant.}
	\label{fig_1}
\end{figure*}
As shown in~\cite{Zhu_Brian_2011,Huang_Ruan_2017}, the surface self-diffusion coefficient $D_s$ can be measured as follows. The surface of an amorphous system should certainly be deformed to form parallel nanogrooves and, then, the mechanical relaxation of this deformed surface should be studied by atomic force and optical microscopy (see Fig.~\ref{fig_1}). Such measurements of the surface self-diffusion coefficient $D_s$ for some various crystallized molecular glasses revealed interesting features. In particular, it was found that the measured surface self-diffusion coefficient $D_s$ correlates with the crystal nucleus growth rate $v_N$~\cite{Huang_Ruan_2017}. Thus, the surface self-diffusion coefficient $D_s$ can be considered in this case as one of the key kinetic characteristics of the nucleation and growth of crystals. Furthermore, it was established in the cited works that the difference between the self-diffusion coefficients in the surface crystallized layer and in the bulk of the system can reach five orders of magnitude and more~\cite{Zhu_Brian_2011,Huang_Ruan_2017}. This difference is due exclusively to the presence of an interface between a high-density molecular system and much less dense air. As a result, local rearrangements of molecules in the surface layer become more pronounced and the effective self-diffusion coefficient becomes much larger than the bulk self-diffusion coefficient. It is noteworthy that this difference in another amorphous system where intermolecular bonds are strong (e.g., in metalic glasses) will be insignificant. Thus, crystallized amorphous systems can also be strongly inhomogeneous in their diffusion and viscous characteristics, which can vary in a wide range. In turn, this can explain the incorrectness of theoretical estimates of the crystal nucleation rate $J_s$, kinetic coefficient (attachment coefficient) $g^+$, and crystal nucleus growth rate $v_N$ obtained in approximations where some kinetic characteristics of crystallization either is directly identified with the self-diffusion coefficient or shear viscosity coefficient, which is determined for \textit{the entire system}, or is expressed in terms of these transport coefficients~\cite{Tropin_2016}. Such approximations are used, e.g., in the Turnbull-Fisher and Kelton-Greer models, as well as in the so-called ballistic model for the coefficient $g^+$~\cite{Mokshin_Galimzyanov_PCCP_2017}. Consequently, it is reasonable to suppose that the results obtained in these approximations will be more accurate if the transport coefficients-viscosity and self-diffusion coefficient-are calculated directly for spatial regions where crystal nuclei are formed and crystallization is initiated.

\begin{figure*}[ht]
	\centering
	\includegraphics[width=0.8\linewidth]{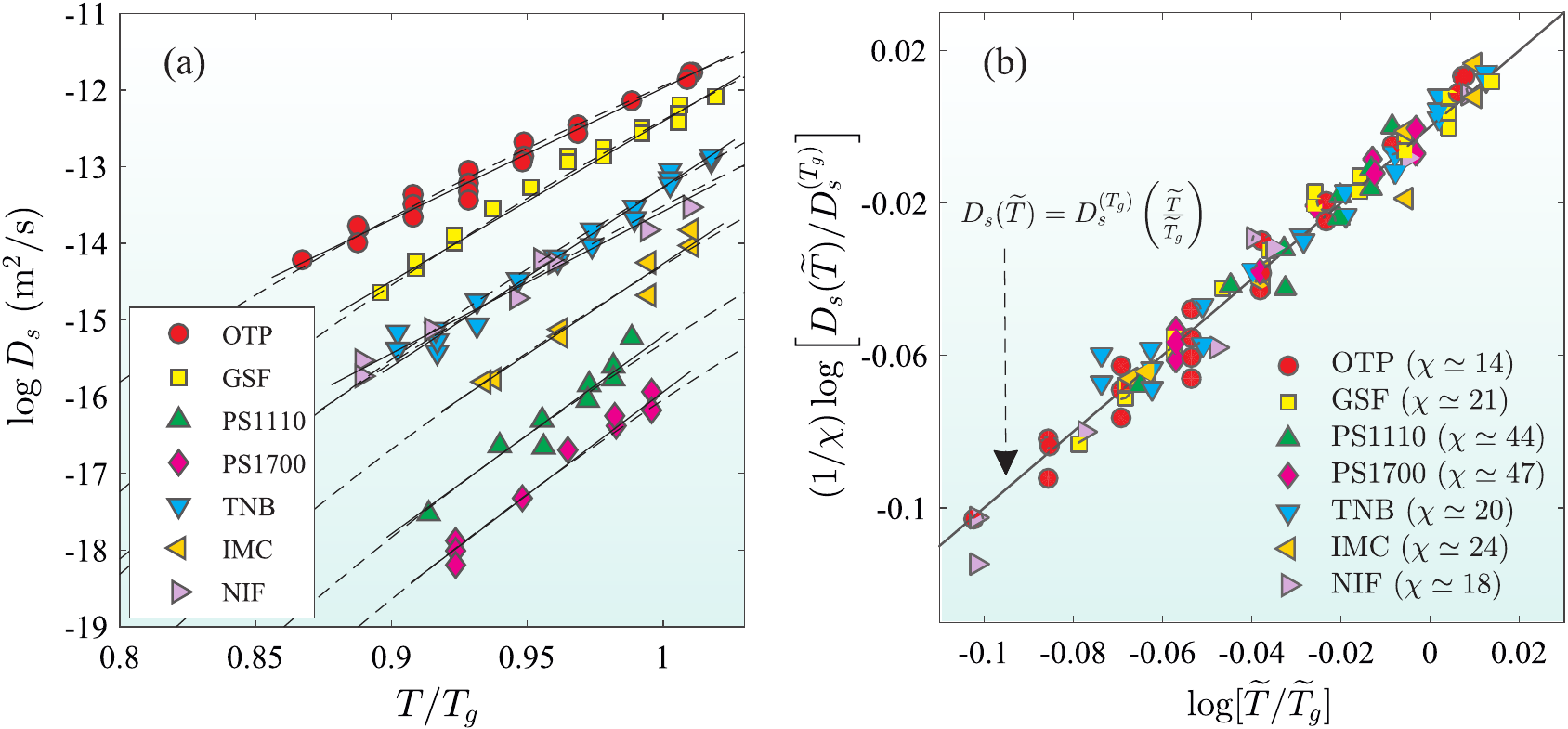}
	\caption{Fig. 2. (Color online) (a) Logarithm of the surface self-diffusion coefficient $D_s$ versus the dimensionless temperature $T/T_{g}$ for various crystallized molecular glasses according to (solid lines) Eq.~(\ref{eq_frenkel}), (dashed lines) Eq.~(\ref{eq_Diff}), and (experimental points)~\cite{Zhang_Brian_2015,Huang_Ruan_2017,Zhang_Yu_2016}. (b) Dimensionless surface self-diffusion coefficient $D_{s}/D_{s}^{(T_{g})}$ versus the dimensionless temperature $\widetilde{T}/\widetilde{T}_{g}$ on a log--log scale. The solid line corresponds to Eq.~(\ref{eq_univ_law}) with $\chi=1$}
	\label{fig_2}
\end{figure*}

The existense of experimental data for the surface self-diffusion coefficient $D_s$ in crystallized molecular glasses reported in~\cite{Zhang_Brian_2015,Huang_Ruan_2017,Zhang_Yu_2016} provides the appropriate possibility of testing the idea of the unified description of temperature dependences of the rate characteristics of crystallization of systems within scaling relations~\cite{Mokshin_Galimzyanov_PCCP_2017,Mokshin_Galimzyanov_2015}. Figure~\ref{fig_2}(a) shows the temperature dependences of the surface self-diffusion coefficient $D_s$ for the following crystallized molecular glasses ~\cite{Zhang_Brian_2015,Huang_Ruan_2017,Zhang_Yu_2016}: \textit{ortho}-terphenyl  (OTP), griseofulvin (GSF), polystyrene oligomers with molar masses of $1110$ and  $1700$~g/mol (PS1110 and PS1700), \textit{tris}-naphthyl benzene (TNB), indomethacin (IMC) and nifedipine (NIF). It is noteworthy that these systems are molecular glasses of different types and have significantly different structures. 
The experimental data for the coefficient $D_s$ shown in Fig.~\ref{fig_2}(a) correspond to temperatures $T\leq T_{g}$, where $T_{g}$ is the glass transition temperature of the system. At such temperatures, a glassy  system is characterized by high viscosity (e.g., shear viscosity) and a very low self-diffusion coefficient of particles. Nevertheless, the experimental method used to measure the coefficient $D_s$ in~\cite{Zhang_Brian_2015} allows the detection of self-diffusion at a level of $10^{-18}$\,m$^{2}$/s. Such self-diffusion coefficients can be expected for thermodynamic states at temperatures comparable with the glass transition temperature $T_g$. According to Fig.~\ref{fig_2}(a), the coefficient $D_s$ for the systems under consideration increases with the temperature, as expected. The experimental data for the coefficient $D_s(T)$ in the considered temperature interval can be reproduced both by the expression
\begin{equation} \label{eq_frenkel}
\lg D_s(T) = \lg D_s^{(0)} + \mathcal{A} T,
\end{equation}
and by the known Eiring-Frenkel model for self-diffusion
\begin{equation} \label{eq_Diff}
\lg D_s(T) = \lg D_s^{(\infty)} - \frac{E}{k_B T} \cdot \lg \mathrm{e}.
\end{equation}
Here, the coefficients $\mathcal{A}$, $D_s^{(0)}$, and $D_s^{(\infty)}$, as well as the activation energy $E$, are independent of the temperature; $k_B$ is the Boltzmann constant, and $\mathrm{e}$ is the Euler constant. Moreover, self-diffusion as a function of the temperature below the melting temperature $T_m$ is no longer a purely activation process and can be reproduced by another model (Vogel-Fulcher-Tammann-Hesse model, mode-coupling theory, etc.). Nevertheless, it can be expected that physical mechanisms of diffusion processes in supercooled liquids and glasses provide a certain unified scenario of change in the character of self-diffusion as a function of the temperature irrespective of a model law or a set of model laws describing the temperature dependence of the coefficient $D_s$ in the entire temperature range $T < T_{m}$. To take into account this unified scenario, it is convenient to use the reduced temperature scale $\widetilde{T}$, where transition (crossover) temperatures such as the melting temperature $T_m$ and glass transition temperature $T_g$ are the same for all systems. In particular, defining the $\widetilde{T}$ scale in the interval  $[0,\;\widetilde{T}_m]$ such that zero temperature is $\widetilde{T}_0=0$, glass transition temperature is $\widetilde{T}_g=0.5$, and melting (liquidus) temperature is $\widetilde{T}_m=1$ for any system, we obtain the following expression for this temperature scale ~\cite{Mokshin_Galimzyanov_2015}:
\begin{equation}\label{eq_scaled_tmp}
\widetilde{T}=K_{1}(T_m,T_g)\left(\frac{T}{T_{g}}\right)+K_{2}(T_m,T_g)\left(\frac{T}{T_{g}}\right)^{2},
\end{equation}
Here, $T$ is the temperature in kelvin, $\widetilde{T}$ is the dimensionless temperature, the glass transition temperature $T_g$ and melting temperature $T_m$ estimated in kelvin for a particular system under consideration are the input parameters, and 
\begin{eqnarray}\label{eq_K1_K2}
K_{1}(T_m,T_g)&=&\frac{0.5T_{m}^{2}-T_g^{2}}{T_{m}(T_{m}-T_g)},\nonumber\\
K_{2}(T_m,T_g)&=&0.5-K_{1}(T_m,T_g). \nonumber
\end{eqnarray}

Figure~\ref{fig_2}(b) shows the scaled surface self-diffusion coefficient $D_{s}/D_{s}^{(T_{g})}$ as a function of the reduced temperature $\widetilde{T}/\widetilde{T}_{g}$. Here, $D_{s}^{(T_{g})}$ is the surface self-diffusion coefficient at the glass transition temperature $T_g$. It is  seen that experimental data are reproduced by the unified power law
\begin{equation}\label{eq_univ_law}
    D_{s}(\widetilde{T})=D_{s}^{(T_{g})}\left(\frac{\widetilde{T}}{\widetilde{T}_{g}}\right)^{\chi},
\end{equation}
where  $\chi > 0$ is the positive exponent depending on the type of the system and on the pressure in the system. The found exponent $\chi$ for the considered crystallized molecular glasses lie in the range from $\chi\simeq14$ (for OTP) to $\chi\simeq47$ (for polystyrene) (see Table $1$). Analysis indicates that a smaller $\chi$ value corresponds to a system with a lower molar mass. Indeed, $\chi\simeq14$ for OTP with a molar mass of of $230$\,g/mol, whereas $\chi\simeq47$ for polystyrene oligomers with a molar mass of $1700$\,g/mol.
\begin{table*}[ht]
	\centering
	\caption{Table 1. Some parameters and characteristics of the considered crystallized molecular systems: glass transition temperature $T_{g}$; melting temperature $T_{m}$; temperature coefficient $K_{1}(T_m,T_g)$ in Eqs.~(\ref{eq_scaled_tmp}) and (\ref{eq_m_chi_2}); surface self-diffusion coefficient $D_{s}^{(T_g)}$ at the glass transition temperature $T_g$; exponent $\chi$ in Eq. (\ref{eq_univ_law}); index of fragility $m$; exponent $\xi$ in the generalized Einstein-Stokes equation and in Eq.~(\ref{eq_ds_law_1}). The values of the parameters $T_{g}$, $T_{m}$, and $m$ are taken from~\cite{Huang_Ruan_2017,Zhang_Yu_2016,Miriglan_Schweizer_2014,Santangelo_1988,Swallen_2011,frag1,frag2,frag3}}\label{tab:1} 
	\begin{tabular}{|c|c|c|c|c|c|c|c|}
		\hline
		System  & $T_g$  & $T_m$  & $K_{1}(T_m,T_g)$ & $D_{s}^{(T_g)}$ & $\chi$ & $m$ & $\xi$ \\
		\hline
		OTP 	& $246$  & $331$ & $-0.204$ & $1.1\times10^{-12}$ & $14\pm3$ & $78 \pm 10$ & $0.187$ \\
		GSF 	& $361$	 & $493$ & $-0.135$ & $4.4\times10^{-13}$ & $21\pm3$ & $73 \pm 12$ & $0.282$ \\
		PS1110	& $307$  & $513$ & $0.353$ 	& $7.1\times10^{-16}$ & $44\pm5$ & $140\pm 15$ & $0.176$ \\
		PS1700 	& $320$  & $533$ & $0.349$ 	& $1.1\times10^{-16}$ & $47\pm5$ & $141\pm 15$ & $0.188$ \\
		TNB  	& $347$  & $467$ & $-0.203$ & $5.2\times10^{-14}$ & $20\pm3$ & $76 \pm 10$ & $0.273$ \\
		IMC  	& $315$  & $431$ & $-0.127$ & $2.1\times10^{-14}$ & $24\pm3$ & $78 \pm 5$  & $0.299$ \\
		NIF  	& $315$  & $446$ & $0.004$ 	& $5.9\times10^{-15}$ & $18\pm2$ & $70 \pm 13$ & $0.221$ \\
		\hline
	\end{tabular}
\end{table*}

The parameter~$\chi$ estimates the rate of variation of the surface self-diffusion coefficient in the unit temperature interval within the temperature range $(0,\;T_m]$.
This estimate is independent of the relation between the melting temperature $T_m$ and the glass transition temperature $T_g$ for a particular system. This means that this parameter can characterize the glass-forming ability of the system and should correspond to some known criterion of the glass-forming ability of the system~\cite{Brazhkin_2019}. Let the surface self-diffusion coefficient $D_s$ be related to the viscosity in the surface layer $\eta_s$ through the generalized Einstein-Stokes relation
\begin{equation} \label{eq_Eins_Stokes}
    D_s= (C\; T/\eta_s)^{\xi}, \hskip 1cm 0 < \xi \leq 1,
\end{equation}
where $C$ is the temperature-independent positive constant with the dimension of pascal multiplied by meter squared per kelvin for $\xi=1$~\cite{Angell_2018}. We recall that the usual Einstein-Stokes relation is no longer valid at temperatures near and below the melting temperature $T_m$ ~\cite{Costigliola_Dyre_2019}. Then, from Eqs.~(\ref{eq_scaled_tmp}), (\ref{eq_univ_law}) and (\ref{eq_Eins_Stokes}) we obtain the following expression for the viscosity:
\begin{eqnarray}\label{eq_ds_law_1}
\eta_s(T) &=& C\; T\left[2K_{1}(T_m,T_g)\left(\frac{T}{T_{g}}\right)\right . \\
& & + \left .  2K_{2}(T_m,T_g)\left(\frac{T}{T_{g}}\right)^{2}\right]^{-\chi/\xi}.  \nonumber
\end{eqnarray}
Using this expression, one can obtain an expression for the index of fragility $m$, which was introduced to classify high-viscosity liquids~\cite{C_Angell} and is defined as
\begin{equation}\label{eq_m}
m=\frac{\partial\lg\eta(T)}{\partial(T_{g}/T)}\Bigg|_{T=T_{g}}.
\end{equation}
The possible indices of fragility $m$ lie in the range $m \in [17,\; 250]$. Small indices $m$ correspond to ``strong'' glass-forming systems, primarily, covalent melts, whereas large indices $m$ correspond to so-called  ``fragile'' liquids with a pronounced non-Arrhenius temperature dependence of he viscosity~\cite{Superfragility}. According to Eqs.~(\ref{eq_ds_law_1}) and (\ref{eq_m}), the expression for $m$ has the form
\begin{equation}\label{eq_m_chi_2}
m = \frac{1}{\ln 10}\left \{ \frac{\chi}{\xi} \left[ 2-  \frac{1 - 2(T_g/T_m)^2}{1- (T_g/T_m)} \right ] - 1 \right \}.
\end{equation}
\begin{figure}[ht]
	\centering
	\includegraphics[width=0.9\linewidth]{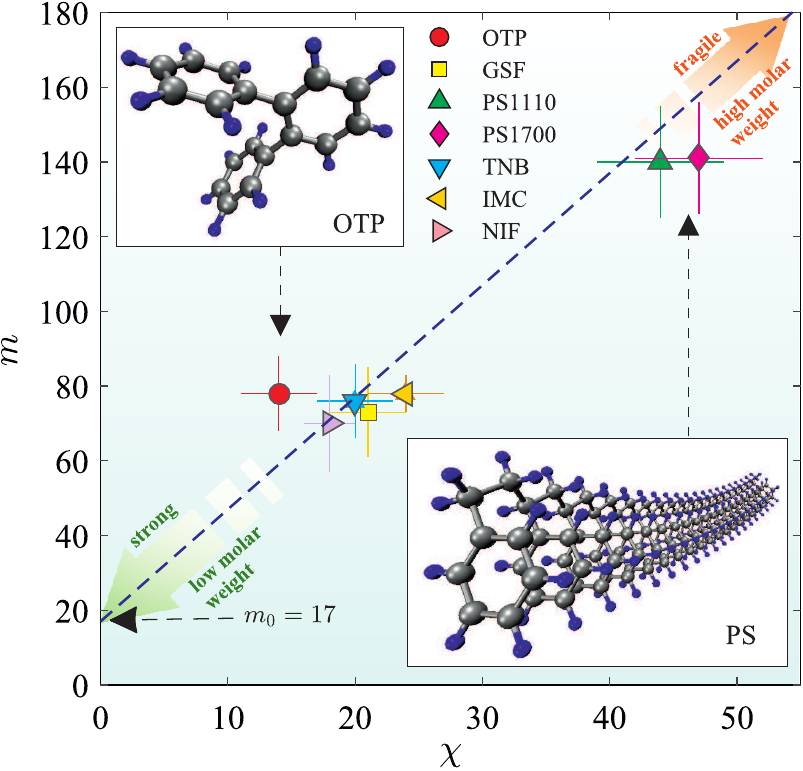}
	\caption{Fig. 3. (Color online) Correlation between the parameter $\chi$ and index of fragility $m$. The dashed line is the reproduction of the correlation by Eq.~(\ref{eq_m_chi_3}). The insets show the configurations of OTP and PS molecules.}
	\label{fig_3}
\end{figure}
This expression relates the following characteristics of the crystallization kinetics of a high-viscosity liquid: the exponent $\chi$ of the general empirical power law for  self-diffusion, which characterizes the mobility of particles; the exponent $\xi$ in the generalized Einstein-Stokes relation, which together with $\chi$ characterizes the viscosity  as a function of the temperature (see Eq.~(\ref{eq_ds_law_1})); the index of fragility $m$ and the ratio $T_{rg} = T_g/T_m$ of the glass transition temperature to the melting temperature, which is considered in~\cite{GFA1,GFA2} as a criterion of the glass-forming ability of a liquid. For the systems considered in this work, all parameters in Eq.~(\ref{eq_m_chi_2}) are known except for the parameter $\xi$.  The estimate of this parameter by means of Eq.~(\ref{eq_m_chi_2}) shows that it lies in the range $0<\xi\leq1$, as expected (see Table 1). According to the existing experimental data, the parameter $\xi$ for most bulk systems in the range from $0.5$ to $0.95$. In particular $\xi>0.79$ for ionic liquids and $\xi>0.67$ for water~\cite{Harris_2009}. Nevertheless, the parameter $\xi$ can be small $0.1<\xi<0.5$, when the diffusivity and viscosity are weakly related to each other. Such a scenario is observed, e.g., when viscous properties of the medium are estimated from the dynamics of the injected molecule with a specific geometry~\cite{Andreozzi_2002,Dzuba_2011} or when the considered system cannot be characterized as bulk, i.e., in the presence of a specific geometry, interfaces, etc. A similar situation with effective self-diffusion and viscosity is considered in this work for the surface layer of molecular systems. This can explain extremely small parameters $\xi$: $0.176\leq\xi\leq0.299$. Finally, correlation between the index of fragility $m$ and the characteristic $\chi$ of the diffusion process is also directly revealed by the direct comparison of their values for different systems. As follows from the $(m,\chi)$ diagram shown in Fig.~\ref{fig_3}, systems classified by the scheme proposed by Angell~\cite{C_Angell} as strong glass-formers are characterized by small parameters $\chi$, whereas brittle glass-forming systems with a high index of fragility $m$ have larger parameters $\chi$. It is remarkable that the found correlation points are located on the  $(m,\chi)$ diagram near the straight line specified by the expression
\begin{equation}\label{eq_m_chi_3}
    m = 3 \chi + m_{0},
\end{equation}
where $m_{0}=17$ is the index of fragility of an ideal ``strong'' glass-forming liquid whose viscosity as a function of the temperature is reproduced by a common Arrhenius dependence both in the equilibrium melt phase and in the supercooled liquid phase~\cite{Kozmidis-Petrovic_2014,Novikov_2016}. It is important that, according to Eq.~(\ref{eq_m_chi_3}), the limit situation with $m=m_0=17$ and $\chi=0$ does not occur because of the absence of systems whose viscosity does not change its character in a wide temperature interval from the temperatures of equilibrium melt to temperatures comparable with $T_g$. In particular, silicon dioxide SiO$_2$ belonging to the ``strongest'' glass-forming systems has $m=19$~\cite{C_Angell}. In addition, it is important that the relations given by Eq. (\ref{eq_m_chi_3}) between the parameters $\chi$ and $m$ shown by the straight line in Fig.~\ref{fig_3} is an approximation obtained in terms of the existing values of these parameters. Expression (\ref{eq_m_chi_2}) specifies a more rigorous relation between these parameters.

To conclude, the results of this work have confirmed that temperature dependences of the kinetic characteristics of crystallization have features universal for different systems and these features can be reproduced by means of universal scaling relations. This has been shown for the surface self-diffusion coefficient, which is directly related to the kinetic coefficient $g^+$ in the case of crystallized molecular glasses. The results can be used (i) to develop the general theory of viscosity of high-density liquids (see, e.g., the discussion in~\cite{Kelton}), (ii) to determine conditions promoting amorphization of liquids, and (iii) to determine optimal physical criteria for estimating the glass-forming ability of liquids.

\section*{Acknowledgement}

We are grateful to Prof. V.N. Ryzhkov (Institute for High Pressure Physics, Russian Academy of Sciences, Troitsk, Moscow) and Acad. V.V. Brazhkin (Institute for High Pressure Physics, Russian Academy of Sciences, Troitsk, Moscow) for valuable advice and discussions of some results of this work. 

\section*{Funding}

This work was supported by the Russian Science Foundation (project no.19-12-00022).

\end{document}